# An Empirical Study of the Impact of Test Strategies on Online Optimization for Ensemble-Learning Defect Prediction

Kensei Hamamoto
*Kindai University*
Higashi-osaka, Japan
2433340478s@kindai.ac.jp

Masateru Tsunoda
*Kindai University*
Higashi-osaka, Japan
tsunoda@info.kindai.ac.jp

Amjed Tahir
*Massey University*
Palmerston North, New Zealand
a.tahir@massey.ac.nz

Kwabena Ebo Bennin
*Wageningen UR*
Wageningen, Netherlands
kwabena.bennin@wur.nl

Akito Monden
*Okayama University*
Okayama, Japan
monden@okayama-u.ac.jp

Koji Toda
*Fukuoka Institute of Technology*
Fukuoka, Japan
toda@fit.ac.jp

Keitaro Nakasai
*OMU College of Technology*
Osaka, Japan
nakasai@omu.ac.jp

Kenichi Matsumoto
*NAIST*
Ikoma, Japan
matumoto@is.naist.jp

*Abstract*— Ensemble learning methods have been used to enhance the reliability of defect prediction models. However, there is an inconclusive stability of a single method attaining the highest accuracy among various software projects. This work aims to improve the performance of ensemble-learning defect prediction among such projects by helping select the highest accuracy ensemble methods. We employ bandit algorithms (BA), an online optimization method, to select the highest-accuracy ensemble method. Each software module is tested sequentially, and bandit algorithms utilize the test outcomes of the modules to evaluate the performance of the ensemble learning methods. The test strategy followed might impact the testing effort and prediction accuracy when applying online optimization. Hence, we analyzed the test order's influence on BA's performance. In our experiment, we used six popular defect prediction datasets, four ensemble learning methods such as bagging, and three test strategies such as testing positive-prediction modules first (PF). Our results show that when BA is applied with PF, the prediction accuracy improved on average, and the number of found defects increased by 7% on a minimum of five out of six datasets (although with a slight increase in the testing effort by about 4% from ordinal ensemble learning). Hence, BA with PF strategy is the most effective to attain the highest prediction accuracy using ensemble methods on various projects.

## I. INTRODUCTION

Owing to the limitations of human resources and development duration, it can be challenging to dedicate substantial resources to extensively test all modules within a project, which can lead to an increase in testing effort. Defect prediction aims to detect defects earlier, which can reduce testing efforts [15][22].

There has been increased attention to using ensemble learning-based methods to improve the prediction accuracy of module-level defect prediction models [13]. Ensemble learning combines the prediction from several models and generates new predictions (e.g., when most models predict a module as defective, the prediction for the module becomes "*defective*").

Although various ensemble-learning methods are available (e.g., *bagging*, *boosting*, and *stacking*), the accuracy of those methods usually varies, depending on the training dataset used [5]. For instance, the accuracy of a model trained on some versions can vary on other versions. This is considered an external validity issue in defect prediction [3]. Therefore, identifying and selecting the accurate method can be challenging.

To help select the ensemble-learning method with the highest accuracy, we apply an online optimization based on bandit algorithms (BA) [8] and evaluate the performance of BA. BA is often explained through an analogy with slot machines. Assume that a player has 100 coins to bet on several slot machines, and the player wants to maximize their reward. BA suggests that the player bets only one coin on each slot machine to seek the best chances. BA seeks sequentially best candidates (referred to as arms) whose expected rewards are unknown to maximize total rewards. For BA to select an ensemble-learning method, we regard slot machines as the methods and playing on a slot machine as testing a module. When the test outcome makes a prediction (i.e., defective or non-defective) on the module, we regard that a coin (reward) is acquired from the arm.

Ensemble-learning methods and BA are similar regarding utilizing multiple prediction models; hence, it is not evident that BA also works well in selecting ensemble methods. Additionally, considering the nature of BA, test strategies such as "testing smaller modules first (SF)" could affect the BA's performance in selecting the highest accuracy ensemble-learning model. For instance, it is probable that the accuracy (i.e., reward) of each ensemble-learning method (i.e., arm) is similar on smaller modules but different on larger modules. When SF is applied, it would be difficult to identify the best (i.e., the highest expected reward) method in the early testing stage. That might affect the accuracy of the prediction obtained by BA. Our study sheds light on such aspects of BA, which previous studies [1][8][19] did not consider. The aims of our study are:

- Help to select ensemble-learning methods to enhance the accuracy of defect prediction.
- Help to select test strategy considering both accuracy and effort when applying BA.

## II. BANDIT ALGORITHM

### A. Procedure

As shown in Figure 1, we assume that module-level defect prediction models are built using ensemble methods, and defects of test target modules are predicted before applying BA. The prediction of each model is treated as an arm. Modules are tested sequentially, and BA selects arms based on the test outcomes of the modules by following the procedure illustrated in Figure 1 [19]. In the procedure, the average reward represents each arm's prediction accuracy (e.g., AUC). For instance, if one intends to select the highest AUC model, AUC is used as the average reward. Below are the steps followed to achieve our goals.

1. Select an arm based on the average reward (AUC) of arms.
2. Test a module based on the prediction of the arm selected in step 1.
3. Recalculate the average reward of each arm, comparing the prediction and the test outcomes (step 2).
4. Return to step 1.

The steps are iteratively performed during testing. Initially, the average reward of all arms is set to zero, and therefore, an arm is selected randomly in the first instance. In step 2, when the defect prediction result is "defective", developers spend more effort on testing. When the results show "non-defective", developers then spend less effort to save resources [22].

For BA's first iteration, from arms A and B, A is selected randomly in step 1. In step 2, module t7 is tested with much testing effort because the prediction result of the selected arm is "defective." In step 3, the prediction of t7 of arm A is evaluated as true-positive, and that of arm B is evaluated as false-negative (i.e., defects are found in t7). Based on this evaluation, we calculate the AUC for each arm. In the second iteration, arm A is still selected in step 1 as the average reward of arm A is higher than B. In step 2, less testing effort for module t5 is spent because the prediction on arm A turns out as "non-defective". In step 3, based on the test outcome, all arms' prediction up to t5 is evaluated as true-negative.

### B. Defect overlooking

**Type 1 overlooking**: When a defect prediction model predicts a negative result (i.e., "non-defective"), developers will typically spend less testing effort writing for those modules to allocate testing resources [15][21] efficiently. As a result, the test overlooks defects, and the module might be regarded as "non-defective," even if it includes defects [19]. This case is called Type 1 overlooking [6].

In Figure 2, the column "test outcome" considers only defects during testing, while "actual outcome after testing" also considers defects after testing is done. In the figure, we assume that defects are overlooked with high probability when the prediction is negative due to less testing effort.

In Figure 2, arm B is randomly selected on the first iteration. The reward of arm B on modules t11 and t19 is true-negative based on the test outcomes. However, based on the actual outcome, this reward is a false-negative. Likewise, based on the test (not actual) outcome, arm A's reward is erroneously set as false-positive. As a result, the AUC value is inaccurate, and arm B, with a low accuracy, is erroneously selected [19].

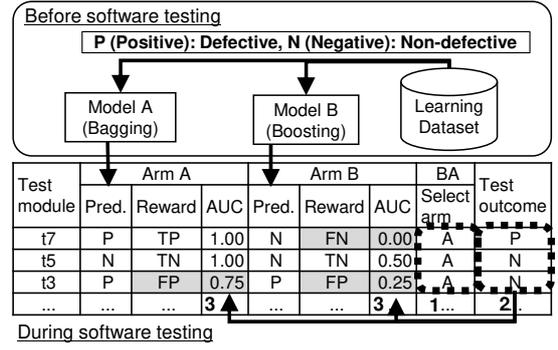

Fig. 1. **Procedure of defect prediction based on BA**

| Test module | Arm A | | | Arm B | | | BA Select arm | Test outcome | Actual outcome after testing |
|---|---|---|---|---|---|---|---|---|---|
| | Pred. | Reward | AUC | Pred. | Reward | AUC | | | |
| t11 | P | **FP** | 0.00 | N | **TN** | 1.00 | B | **N** | P |
| t19 | P | **FP** | 0.00 | N | **TN** | 1.00 | B | **N** | P |
| t15 | N | TN | 0.33 | N | TN | 1.00 | B | N | N |
| t13 | P | **FP** | 0.20 | P | **FP** | 0.80 | B | N | P |
| ... | | | | | | | | | |

**Type 1**: Occur in most cases    **Type 2**: Occur about 20% probability

Fig. 2. **Type 1 and Type 2 defect overlooking**

| Test module | Arm A | | | Arm B | | | BANP | | Test outcome | Actual outcome after testing |
|---|---|---|---|---|---|---|---|---|---|---|
| | Pred. | Reward | AUC | Pred. | Reward | AUC | Select arm | Pred. | | |
| t11 | P | TP | 1.00 | N | FN | 0.00 | -- | **P** | P | P |
| t19 | P | TP | 1.00 | N | FN | 0.00 | -- | **P** | P | P |
| t15 | N | TN | 1.00 | N | TN | 0.50 | A | - | N | P |
| t13 | P | **FP** | 0.75 | P | **FP** | 0.25 | A | - | N | P |
| ... | | | | | | | | | | |

Occur about 20% probability

Fig. 3. **Procedure of BANP**

**Type 2 overlooking**: Even when the prediction is positive (i.e., "defective"), defects are sometimes overlooked during testing [1][19]. This case is called Type 2 overlooking [6]. This could occur even when a defect prediction result properly informs the allocation of testing resources (i.e., extensive resources). Module t13 (shown in Figure 2) is an example of such a case. Based on large-scale data from cross-companies, about 17% of defects are overlooked during integration testing [9].

**Handling Type 1 overlooking with BA**: To suppress the influence of Type 1 overlooking, Tsunoda et al. proposed BANP (Bandit Algorithm to handle Negative Prediction) [19]. BANP regards prediction as positive during BA's early iteration. In summary, BANP forcibly changed the prediction on about 10% of the modules.

Figure 3 illustrates how BANP works. BANP sets "P" on the "Pred.-BANP" column for modules t11 and t19 (i.e., early iterations). As a result, t11 and t19 are regarded as positive-prediction modules, and Type 1 overlooking is suppressed by testing because much testing effort is spent on the module. Although the "defect overlook" does occur on modules t11 and t19 in Figure 2, these modules are rewarded, and accurate AUC values are obtained in Figure 3. As a result, the proper arm (i.e., arm A in Figures 2 and 3) is selected. That enhances the accuracy of BA.

| Test module | LOC | Test effort | Test-effort ratio | Arm B | | Test result | Actual outcome after testing | Probability of Type 1 |
|---|---|---|---|---|---|---|---|---|
| | | | | Pred. | Reward | | | |
| t56 | 1000 | 10.00 | 100% | P | TP | P | P | 0% |
| t55 | 1000 | 2.50 | **25%** | N | **TN** | **N** | P | **75%** |

Fig. 4. **Relationship between test-effot ratio and probability of Type 1 ovelooking**

| Test module | LOC | Test effort | Arm A | | AUC | BANP | | Test outcome |
|---|---|---|---|---|---|---|---|---|
| | | | Pred. | Reward | | Select arm | Pred. | |
| t21 | 3300 | **33.00** | N | TN | 1.00 | -- | P | **N** |
| t29 | 2900 | **29.00** | N | TN | 1.00 | -- | P | **N** |
| Sum of excessive effort = 62.00 | | | ... | ... | ... | ... | ... | ... |

(a) Applying LF strategy

| Test module | LOC | Test effort | Arm A | | AUC | BANP | | Test outcome |
|---|---|---|---|---|---|---|---|---|
| | | | Pred. | Reward | | Select arm | Pred. | |
| t31 | 180 | **1.80** | N | TN | 1.00 | -- | P | **N** |
| t35 | 200 | **2.00** | N | TN | 1.00 | -- | P | **N** |
| Sum of excessive effort = 3.80 | | | ... | ... | ... | ... | ... | ... |

(b) Applying SF strategy

Fig. 5. **Testing effort of BANP**

| Test module | LOC | Arm A | | | Arm B | | | BANP | | Test outcome |
|---|---|---|---|---|---|---|---|---|---|---|
| | | Pred. | Reward | AUC | Pred. | Reward | AUC | Select arm | Pred. | |
| t48 | 2742 | **P** | TP | 1.00 | N | FN | 0.00 | - | P | P |
| t44 | 1964 | N | TN | 0.75 | **P** | TP | 0.50 | - | P | P |
| t49 | 1523 | **P** | TP | 0.90 | N | FN | 0.33 | - | P | P |
| | | | | | | | | ... | ... | ... |

Predicted "defective" by arms — Sort by size

Fig. 6. **Example of PF strategy**

## C. Testing effort

**Definition**: Testing effort is known to increase as module size increases [11]. Additionally, as explained in Section B, testing resources (i.e., effort) vary depending on the prediction results. We define the ratio of testing effort on negative to positive-prediction modules as the **test-effort ratio**. Based on the assumptions, we regarded the testing effort as follows:

$$effort = \begin{cases} size \cdot c & if\ pred = 1, \\ size \cdot c \cdot ratio & if\ pred = 0 \end{cases} \quad (1)$$

In the equation, *size* signifies module size, such as LOC (lines of code), *c* is constant, and *pred* denotes prediction results (1: positive, 0: negative). For instance, Figure 4 shows a case where $c = 0.01$ and the test-effort ratio = 0.1.

**Probability of Type 1 overlooking**: If the ratio is 1.0, the effort of negative prediction modules is the same as positive ones, and the probability of Type 1 overlooking is 0%. When the ratio gets smaller, the probability conversely gets larger. Therefore, we assumed a proportional relationship between the effort and the probability, and when the ratio is *n*%, the probability is 1 - *n*%. For instance, as shown in Figure 4, when the ratio is 25%, the probability is 75% on module t55.

## D. Test strategy

One of the significant test approaches is risk-based testing [10]. In software testing, this approach prioritizes risky modules that could have high-probability defects or include many functions (i.e., larger-size modules). When we focus on such aspects, a strategy for testing larger modules first (LF) is reasonable because they would have more functions and potentially more defects with an established relationship between the size and the number of defects [17].

However, when applying BANP, LF could increase testing efforts. In Figure 5 (a), the test-effort ratio is 0.1, and the effort for modules t21 and t29 is large because their size is also large. However, they are non-defective modules. That is, the testing efforts for the modules are regarded as excessive (i.e., great effort should not be allocated). If smaller modules are tested first (**SF**), such excessive effort becomes smaller, even if the prediction by BANP is incorrect, as shown in Figure 5 (b).

Additionally, as shown in Figure 6, after sorting by their module size, positive-prediction modules are tested first (**PF**), which is also reasonable, considering both effort and quality. This is because PF could avoid test modules with low probability defects. A similar strategy is mentioned in [1][20]. In the figure, modules t48, t44, and t49 are predicted as defective by one of the arms, and the modules are sorted by size.

## III. EXPERIMENT

### A. Setup

**Dataset**: We used six projects from the NASA [7] and PROMISE [2] repositories, which have been widely used in ensemble-learning studies [12][14][16][18]. The NASA dataset contains a set of metrics and defect data collected from several NASA projects. The PROMISE defect data was collected from open-source projects. We selected three datasets from each repository, considering the diversity of each dataset's size and the ratio of defective modules. Table 1 shows details of the datasets used in the experiment.

TABLE I. USED DATASETS

(a) Number of modules on Promise repository

| Software | Ver. | Learning dataset | | Ver. | Test dataset | |
|---|---|---|---|---|---|---|
| | | All | Defective | | All | Defective |
| ant | 1.6 | 351 | 92 (26.2%) | 1.7 | 745 | 166 (22.3%) |
| prop | 5 | 8516 | 1299 (15.3%) | 6 | 660 | 66 (10.0%) |
| synapse | 1.1 | 222 | 60 (27.0%) | 1.2 | 256 | 86 (33.6%) |

(b) Number of modules on NASA repository

| Project | All | Defective |
|---|---|---|
| KC4 | 125 | 61 (48.8%) |
| MW1 | 403 | 31 (7.7%) |
| PC4 | 1458 | 178 (12.2%) |

**Evaluation criteria**: We used AUC to evaluate the accuracy of the prediction models- a widely used metric in previous defect prediction studies [12][14][16][18]. The maximum value of AUC is 1. When the value of a prediction model is large, it means that the model's prediction accuracy is high. We only used AUC as a criterion because BA optimizes defect prediction based on AUC, as explained in Section II.A.

We also used testing effort as one of the evaluation criteria. In formula (1), we set the constant *c* to 1 because the constant can be omitted when comparing the effort among methods.

We also defined **RDIFF** (relative difference) to compare a criterion as follows:

$$RDIFF = 1 - \frac{target}{baseline} \quad (2)$$

TABLE II. AVERAGE AUC OF TEST STRATEGIES ON SIX DATASETS

| Strategy | SF | | | LF | | | PF | | |
|---|---|---|---|---|---|---|---|---|---|
| Test-effort ratio | 0.1 | 0.25 | 0.5 | 0.1 | 0.25 | 0.5 | 0.1 | 0.25 | 0.5 |
| $\varepsilon = 0$ | 46 (0.638) | 45 (0.640) | 31 (0.646) | 26 (0.651) | 11 (0.655) | 19 (0.653) | 1 (0.659) | 2 (0.658) | 20 (0.653) |
| $\varepsilon = 0.1$ | 44 (0.640) | 38 (0.643) | 34 (0.645) | 25 (0.651) | 18 (0.653) | 6 (0.656) | 4 (0.656) | 14 (0.654) | 10 (0.655) |
| $\varepsilon = 0.2$ | 42 (0.640) | 41 (0.641) | 33 (0.646) | 23 (0.652) | 27 (0.650) | 24 (0.652) | 13 (0.655) | 16 (0.654) | 12 (0.655) |
| $\varepsilon = 0.3$ | 39 (0.642) | 37 (0.644) | 32 (0.646) | 29 (0.650) | 30 (0.649) | 21 (0.653) | 5 (0.656) | 9 (0.655) | 15 (0.654) |
| UCB | 43 (0.640) | 35 (0.644) | 36 (0.644) | 28 (0.650) | 22 (0.652) | 17 (0.654) | 3 (0.657) | 7 (0.656) | 8 (0.655) |

TABLE III. RDIFF OF TESTING EFFORT (BASELINE: SF)

| Strategy | LF | | | PF | | |
|---|---|---|---|---|---|---|
| Test-effort ratio | 0.1 | 0.25 | 0.5 | 0.1 | 0.25 | 0.5 |
| RDIFF (%) | 4.8 | 2.5 | 0.8 | 12.2 | 5.7 | 2.1 |

TABLE IV. AUC AND ITS RANK OF EACH PREDICTION METHOD

| | ant | KC4 | MW1 | PC4 | prop | synapse | Avg. AUC | Avg. rank |
|---|---|---|---|---|---|---|---|---|
| BA: $\varepsilon = 0$ | 2 (0.700) | 2 (0.825) | 2 (0.584) | 2 (0.731) | 4 (0.498) | 4 (0.614) | 1 (0.659) | 2.7 |
| BA: UCB | 1 (0.700) | 3 (0.822) | 2 (0.584) | 3 (0.723) | 5 (0.498) | 3 (0.615) | 2 (0.657) | 2.8 |
| Bagging | 6 (0.662) | 1 (0.829) | 4 (0.546) | 6 (0.679) | 2 (0.500) | 5 (0.601) | 5 (0.636) | 4.0 |
| RF | 4 (0.694) | 4 (0.802) | 4 (0.546) | 4 (0.689) | 6 (0.493) | 2 (0.616) | 4 (0.640) | 4.0 |
| Stacking | 3 (0.696) | 6 (0.752) | 6 (0.542) | 5 (0.683) | 2 (0.500) | 1 (0.619) | 6 (0.632) | 3.8 |
| XGBoost | 5 (0.686) | 5 (0.779) | 1 (0.588) | 1 (0.780) | 1 (0.508) | 6 (0.589) | 3 (0.655) | 3.2 |

For instance, when comparing the effort of LF and SF, the effort of SF is set to *baseline*, and that of LF is set to *target*.

**Prediction models**: We employed four widely known ensemble-learning methods: *bagging*, *XGBoost*, *random forest (RF)*, and *stacking*. To perform stacking, we made prediction models using linear discriminant analysis, random forest, and generalized boosted models and merged the prediction results using random forest. The method is sometimes called *blending* [5].

We used two BA methods, ε-greedy and UCB method. We set the parameter ε as 0, 0.1, 0.2, and 0.3 because the values are often used to set ε [21].

**Procedure**: When we used the NASA datasets, we applied the hold-out method to evaluate the prediction accuracy. The dataset was randomly separated into learning and testing sets, and the ratio of learning to testing sets size is 3:1. With the PROMISE repository, we selected datasets collected from different software versions to perform cross-version defect prediction. When evaluating BA, we applied it 20 times on each dataset and calculated the average AUC from the 20 repetitions. This is because ε-greedy randomly selects one of the arms (i.e., prediction models) with a probability of 1 - ε; hence, we tried to align the influence of the randomness.

To analyze the influence of the test-effort ratio on defect prediction, we set the ratio at 10%, 25%, and 50%. Therefore, the probability of Type 1 overlooking was 90%, 75%, and 50%, respectively, as explained in Section II.C. We set the probability of Type 2 overlooking at 20% and artificially turned the reward as inaccurate at 20% because about 17% of defects are overlooked during integration testing [9]. When an ensemble-learning method is randomly selected from candidates of methods, the expected performance is the average of the candidates. For instance, the expected AUC is the average AUC of the four methods. We set such an average as the benchmark, and when the performance of BA was higher than that, BA was regarded as effective.

*B. Result*

**Influence of test strategy on AUC**: We analyzed the difference in prediction accuracy among test strategies on BA. Table 2 shows AUC and their overall rank on the three strategies and three patterns of test-effort ratio. In the table, AUC is shown in parentheses. Due to page limitations, we only present the average AUC from six datasets. For a better visual display of the rank, we used shading in the table (darker cells represent higher ranks). We included the benchmark (i.e., the average AUC of four ensemble-learning methods) in the ranking. The rank of the benchmark was 40.

As shown in Table 2, SF was ranked lower, and PF was higher. When the test-effort ratio was 0.1, SF's ranks were lower than the benchmark except for ε = 0.3. Therefore, BA should not be used when SF is applied, and the test-effort ratio is 0.1. Meanwhile, when PF was applied, and the test-effort ratio was 0.1, the rank was higher than the benchmark and most other ratios. Therefore, when we prioritize prediction accuracy, PF can be applied.

**Influence of test strategy on effort**: To analyze the relative difference in testing effort among strategies, we calculated *RDIFF*, as shown in Table III. The table shows the average *RDIFF*, stratifying by test-effort ratio. (we limit this to a test-effort ratio due to space limitations). As shown in the table, when the test-effort ratio was larger, *RDIFF* was smaller. *RDIFF* of PF was larger than LF, and PF increased the effort by 12.2% when the test effort ratio was 0.1. As explained later, there was a positive relationship between prediction accuracy and effort. Hence, considering the accuracy, the larger testing effort is not a severe drawback of PF.

**Comparison of AUC of BA with ensemble learning**: We compared the prediction accuracy of BA with ensemble learning. Based on Table II, we picked up ε-greedy (ε = 0) and UCB when test-effort ratio = 0.1 as representatives of BA methods. Table IV shows each method's AUC and rank on each dataset. Unlike Table II, the rank in Table IV was settled on each dataset. The average rank across the datasets is shown on the rightmost columns. As shown in the table, the average AUC and rank of ε-greedy were the highest among the methods. UCB was the second highest. Additionally, the rank of ε-greedy was higher than the two ensemble methods, at least on each dataset (except for the prop dataset). This suggests that the accuracy of ε-greedy is stable among datasets, and our approach mitigates the external validity issue in defect prediction compared with the conventional ensemble-learning approach.

**Comparison of BA effort with ensemble learning**: There was a positive relationship between testing effort and AUC and between AUC and the number of positive predictions. The correlation coefficient of the former one was 0.76, and the latter one was 0.90. More testing effort is assigned to modules predicted as positive, as explained in Section II.B. As a result, more significant effort is assigned when there are more positive-prediction modules. This is why there is a positive relationship between AUC and effort.

Based on the observation, instead of testing the benchmark's effort (i.e., the average effort among ensemble-learning methods), we picked up that of XGBoost, which is the highest accuracy among ensemble-learning methods. Table V shows *RDIFF*, setting the effort of XGBoost as the baseline. In the table, the average *RDIFF* was 0.9%, and the median was 4% (i.e., BA increased the effort by 4% on average).

*C. Threat to the validity*

**Internal validity**: In Table IV, the difference of AUC between greedy ($\varepsilon = 0$) with XGBoost was 0.004. Such a difference is considered marginal. As shown in Table 4, most of the modules are non-defective; hence, AUC did not reflect the difference in the true-positive rate. Table VI shows the number of true positive (i.e., found defects by prediction methods) and *RDIFF* of BA ($\varepsilon$-greedy), setting the average ones of ensemble methods (i.e., the benchmark) as the baseline.

In Table VI, although the average RDIFF was 1%, this was affected by the difference in the prop dataset. Most ensemble learning did not find any defects that affected the performance of BA. Except for the prop dataset, RDIFF was at 7% at minimum, and the average was 10.2%. The result indicates that BA can find more defects, thus improving software quality. Therefore, considering the influence of BA on effort (i.e., 4% on average), BA is expected to be effective in optimizing ensemble-learning defect prediction.

**External validity**: We used two datasets in our experiment, containing data from open-source and proprietary software. The open-source datasets were used for cross-version defect prediction. It is possible that the ratio of defective modules could affect the performance of BA and defect prediction. Considering that, we selected the MW1 and prop datasets, which both include fewer defective modules, and the synapse and KC4 datasets, which both include a higher number of defective modules. Given the datasets' diversity, we believe this has minimized the threats to external validity.

## IV. RELATED WORK

**Ensemble-learning**: Several studies used ensemble learning for defect prediction [12][14][16][17]. Matloob et al. [13] conducted a systematic literature review of ensemble learning on software defect prediction and clarified the studies' tendencies, such as frequently used techniques (e.g., random forest, boosting, and bagging) and evaluation criteria, such as AUC. However, to our knowledge, no study has applied online optimization to ensemble learning on software defect prediction.

**Bandit algorithms**: BA has been applied to optimize software defect prediction. For instance, Asano et al. [1] used BA to optimize learning data on cross-project defect prediction.

TABLE V. *RDIFF* (%) OF TESTING EFFORT OF E-GREEDY (BA) (BASELINE: XGBOOST)

| ant | KC4 | MW1 | PC4 | prop | synapse | Avg. | Median |
|-----|-----|-----|-----|------|---------|------|--------|
| 7.4 | 2.6 | 5.4 | 2.6 | -24.7 | 12.5 | 0.9 | 4.0 |

TABLE VI. *RDIFF* OF FOUND DEFECTS OF E-GREEDY (BA) (BASELINE: AVERAGE OF ENSEMBLE-LEARNING)

| Dataset (Actual defects) | Bagging | RF | Stacking | XGBoost | Avg. Ens. | BA | *RDIFF* (%) |
|---|---|---|---|---|---|---|---|
| ant (166) | 66 | 81 | 82 | 81 | 77.5 | 85.2 | 9.9 |
| KC4 (20) | 16 | 14 | 12 | 14 | 14.0 | 16.3 | 16.1 |
| MW1 (10) | 1 | 1 | 1 | 2 | 1.3 | 2.0 | 60.0 |
| PC4 (59) | 23 | 23 | 22 | 35 | 25.8 | 27.6 | 7.0 |
| prop (66) | 0 | 0 | 0 | 3 | 0.8 | 0 | -96.7 |
| Synapse (166) | 24 | 29 | 26 | 26 | 26.3 | 28.7 | 9.1 |

Tsunoda et al. [19] applied BA to select feature reduction techniques. Similarly, Hayakawa et al. [8] used BA to optimize prediction methods such as logistic regression and decision trees and compared the accuracy of BA to majority voting. However, past studies did not treat the optimization of ensemble learning, testing effort, test-effort ratio, and testing strategies.

Both BA and ensemble learning use several models to make their predictions. However, BA is an online optimization approach that utilizes test outcomes during the testing phase. In contrast, ensemble learning is not an online optimization or learning approach and does not utilize test outcomes.

## V. CONCLUSION

We apply and analyze the performance of online optimization based on Bandit Algorithms (BA) to ensemble learning defect prediction to attain stable prediction accuracy of ensemble learning on various datasets. BA compares actual test outcomes (i.e., defective or not) of modules with the prediction of ensemble learning methods to evaluate the accuracy of each method. The evaluation sequence depends on the test strategy followed (i.e., which modules are tested preferentially), which could, in turn, affect the prediction accuracy and testing effort. We analyzed the impact of three test strategies (i.e., LF: testing larger modules first, SF: testing smaller modules first, and PF: testing positive-prediction modules first) on BA. In the experiment, we changed the test-effort ratio (i.e., the ratio of testing effort on negative to positive-prediction modules). Our findings show that:

- The test strategy affects both BA's prediction accuracy and testing effort. PA showed the highest accuracy but required the most extensive testing effort.
- Even when the test-effort ratio was set to 0.1, PA still showed the highest accuracy.
- BA steadily improved the prediction accuracy of ensemble methods but slightly increased testing efforts.

Our result suggests that BA with PA is the best approach to enhance the prediction accuracy of online learning, and setting a test-effort ratio of 0.1 is sufficient to achieve high prediction accuracy. Those findings are preliminary. To enhance the reliability of the results, we plan to conduct experiments that consider modules' complexity (alongside size) with various datasets in the future.